\documentclass[aps,pra,onecolumn,groupedaddress,floatfix,longbibliography,superscriptaddress,notitlepage]{revtex4-1}
\usepackage{amsmath}
\usepackage{amssymb}
\usepackage{amsfonts}
\usepackage{graphicx}
\usepackage{bbold}
\usepackage{bm}
\usepackage{bm}
\usepackage{cancel}
\usepackage{times,float}
\usepackage{graphicx}
\usepackage[usenames,dvipsnames,svgnames]{xcolor}
\usepackage{hyperref}
\hypersetup{colorlinks=true, linkcolor=NavyBlue, citecolor=PineGreen,urlcolor=cyan}
\usepackage{multirow}
\usepackage{braket}
\usepackage{float}
\usepackage{diagbox}
\usepackage{epsfig,subfig}
\usepackage{tabularx}
\usepackage{changepage}
\usepackage{soul}

\begin{document}

\title{Exact classical limit of the quantum bouncer}

\author{Juan A. Ca\~{n}as}
\affiliation{Universidad Ju\'{a}rez Aut\'{o}noma de Tabasco, Divisi\'{o}n Acad\'{e}mica de Ciencias B\'{a}sicas, 86690 Cunduac\'{a}n, Tabasco, M\'{e}xico}

\author{J. Bernal}
\affiliation{Universidad Ju\'{a}rez Aut\'{o}noma de Tabasco, Divisi\'{o}n Acad\'{e}mica de Ciencias B\'{a}sicas, 86690 Cunduac\'{a}n, Tabasco, M\'{e}xico}

\author{A. Mart\'in-Ruiz}
\email{alberto.martin@nucleares.unam.mx}
\affiliation{Instituto de Ciencias Nucleares, Universidad Nacional Aut\'{o}noma de M\'{e}xico, 04510 Ciudad de M\'{e}xico, M\'{e}xico}

\begin{abstract}
In this paper we develop a systematic approach to determine the classical limit of periodic quantum systems and applied it successfully to the problem of the quantum bouncer. It is well known that, for periodic systems, the classical probability density does not follow the quantum probability density. Instead, it follows the local average in the limit of large quantum numbers. Guided by this fact, and expressing both the classical and quantum probability densities as Fourier expansions, here we show that local averaging implies that the Fourier coefficients approach each other in the limit of large quantum numbers. The leading term in the quantum Fourier coefficient yields the exact classical limit, but subdominant terms also emerge, which we may interpret as quantum corrections at the macroscopic level. We apply this theory to the problem of a particle bouncing under the gravity field and show that the classical probability density is exactly recovered from the quantum distribution. We show that for realistic systems, the quantum corrections are strongly suppressed (by a factor of $\sim 10^{-10}$) with respect to the classical result.
\end{abstract}

\maketitle

\section{Introduction} \label{Intro}

The correspondence principle states that any new theory in physics should reduce, in the appropriate limit, to preceding theories that have been proven to be valid. Besides, the new theory should also explain new phenomena beyond the validity of the older theory. For example, special relativity reduces to classical mechanics for velocities small as compared with the velocity of light in vacuum. Similarly, general relativity reduces to Newtonian gravitation in the limit of weak gravitational fields. The quantum-to-classical correspondence is much more subtle, so much that after a century of the inception of the quantum theory it remains problematic. Put simply, it refers to that any model describing the behaviour of quantum systems must yield the same results as classical physics in the macroscopic limit. 

Although its conceptual and technical importance is largely undisputed, still there is no agreement concerning how the quantum-to-classical correspondence should be defined, and the main difficulty is perhaps that the conceptual framework of both theories are quite different. In classical mechanics we deal with quantities that are directly observable, such as the position and momentum, which indeed define the classical trajectory of the particle. In quantum mechanics, however, we are concerned with the probability density, which specifies the probability  of finding the particle at some position or bearing some momentum.  By many other aspects like this, the quantum-to-classical transition remains a difficult task.

There are two different formulations of the correspondence principle between quantum and classical physics \cite{Makowski_2006}.  One is due to Planck, who postulated that the quantum theory reduces to the classical theory in the limit $h \to 0$. The other, due to Bohr, asserts that the classical behaviour emerges from the quantum world for large energy quantum number $n$. As shown by Liboff, when applied to transition frequencies, these two formulations of the correspondence principle are not universally equivalent  \cite{Liboff_1984}. A combination of both has also been proposed to characterize the classical limit: taking simultaneously $h \to 0$ and $n \to \infty$ subject to the constraint that $J=n h $ is finite, where $J$ is the appropriate classical action \cite{Hassoun_Kobe_1989}. There are a few proposals, all of them rooted in the Planck and Bohr correspondence principles, addressing the intricate connection between the classical and quantum worlds. For example, as Wentzel, Kramers and Brillouin found, an approximate solution to the Schr\"{o}dinger equation (for slowly varying potentials) yields, in the limit $h \to 0$, to the Hamilton-Jacobi equation of classical physics \cite{Wentzel_1926, Kramers_1926, Brillouin_1926}.  Planck's limit has also been used to characterize the classical limit of the Feynman path integral formulation of quantum mechanics \cite{FeynmanHibbs_1965}.  As we shall discuss later, neither the WKB approximation nor the Lagrangian formulation yields the complete classical limit when $h \to 0$.  An additional form of the quantum-classical correspondence is given by Ehrenfest's theorem \cite{Ehrenfest_1927}, which states that  the quantum-mechanical averages of the position and momentum operators obey the classical equations of motion. However, as discussed in Ref. \cite{Ballentine_1994},  this is not true at all if the width of the probability distribution in position space is not negligible.

In a less sophisticated level,  many textbooks \cite{Liboff_Book, Robinett_2006} and articles \cite{Robinett_1995, Doncheski_2000, Robinett_1996, Robinett_2002, Yoder_2006, Semay_2016} compare the quantum mechanical and classical probability densities and conclude, by inspection, that they approach each other in a locally averaged sense in the limit of large energy quantum number $n$.  Nevertheless, it has been argued that this comparison is unsuitable when applied to a pure energy eigenstate, since it is an inherently oscillating state, rather it should be applied to a mixed ensemble of quantum coherent states \cite{Leubner_1988}. The local averaging required to show that the oscillatory quantum probability density approaches the smooth classical distribution is technically quite difficult. It can actually be done analytically only for a particle in a box because of the simplicity of the involved quantum and classical distributions  \cite{Robinett_1995}, but its application to more intricate periodic quantum systems is still lacking.  Our main goal in this paper is reformulate the local averaging process in a simpler and manageable fashion.  The method is based in some simple assumptions, such as the periodicity of the classical motion, the highly oscillatory behaviour of the quantum distribution for large energy quantum number, and that they converge each other when averaged locally over a finite interval. This allow us to express the quantum-classical correspondence in the Fourier space as follows: in the limit of large energy quantum number $n$, the quantum Fourier coefficient asymptotically behaves as the classical Fourier coefficient.  A series of subleading terms (in inverse powers of $n$) may emerge, which can be interpreted as quantum corrections at the macroscopic level.  This method has been successfully applied to the infinite square well potential, the harmonic oscillator and the hydrogen atom \cite{ClassLim1,ClassLim3,ClassLim2}. Here we shall apply this theory to the quantum bouncer and show that the classical probability density is exactly recovered from the quantum distribution. Our results suggest that quantum mechanics does not converge to classical mechanics, but that in the classical limit the distinctive quantum phenomena become strongly suppressed, leaving us with an apparent world described consistently by classical laws.  We conclude also that the local averaging procedure can be applied to one-particle energy eigenstates,  thus refuting the critique of Ref. \cite{Leubner_1988}. 

This paper is organized as follows. In Sec. \ref{Formulation} we introduce the general formulation of the classical limit. The exact classical limit of the quantum bouncer in obtained in Sec. \ref{ClassicalLimit-QB}.  Finally, in Sec. \ref{Conclusions} we discuss our results and suggest further avenues of study.

\section{General formulation of the classical limit} \label{Formulation}

The quantum-classical correspondence is subtle and still not fully understood. It involves both conceptual and formal aspects. Therefore, for establishing a theoretical formalism concerning the classical limit of quantum mechanics, one must bring both theories into a common mathematical framework. As discussed in Sec. \ref{Intro},  very often, a comparison between the quantum and classical  probability densities for periodic systems is used to illustrate how Bohr's correspondence principle works 
\cite{Liboff_Book, Robinett_2006,Robinett_1995, Doncheski_2000, Robinett_1996, Robinett_2002, Yoder_2006, Semay_2016} While the former is defined by $\rho ^{\mbox{\scriptsize qm}} _{n} (x) = \vert \psi _{n} (x) \vert ^{2}$, where $\psi _{n}$ is an energy eigenstate of the system, the latter is properly defined by $\rho _{\mbox{\scriptsize cl}} (x) = \frac{2}{\tau \vert v (x) \vert }$, where $\tau$ and $v (x)$ are the classical period and the local speed, respectively. The comparison between these is not satisfactory at all, since in the limit of large quantum number $n$, the highly oscillatory quantum distribution $\rho ^{\mbox{\scriptsize qm}} _{n} (x)$ does not converge pointwise to the smooth classical probability density $\rho _{\mbox{\scriptsize cl}} (x)$. Instead, they approach each other in a locally averaged sense, i.e.
\begin{align}
\rho _{\mbox{\scriptsize cl}} (x) = \lim _{n \gg 1} \frac{1}{2 \epsilon _{n}} \int _{x - \epsilon _{n}} ^{x + \epsilon _{n}} \rho ^{\mbox{\scriptsize qm}} _{n} (y) \, dy , \label{LocalAverage}
\end{align}
where the interval $\epsilon _{n}$ decreases with increasing the quantum number $n$. The local averaging (\ref{LocalAverage}) can be analytically evaluated only for the most familiar of all quantum mechanical bound state problems: the infinite well. However, it is impractical for more intricate bound-state systems for which wave functions involve special functions (e.g. the harmonic oscillator, the quantum bouncer and the hydrogen atom), and have at best can be treated numerically. The main goal of this section is reformulate the local averaging process of Eq. (\ref{LocalAverage}) in a simpler and manageable fashion. The method is based on the following simple and well-known assumptions:
\begin{itemize}

\item[(i)] the classical motion is periodic, i.e. the particle bounces back and forth between the turning points,

\item[(ii)] the quantum probability density rapidly oscillates in the limit of large energy quantum number $n$, 

\item[(iii)] for large $n$, the classical and quantum distributions approach each other when averaged locally over a finite interval. 

\end{itemize}

Assumptions (i) and (ii) imply that both the quantum and classical probability densities can be expressed as Fourier expansions:
\begin{align}
\rho _{\mbox{\scriptsize cl}} (x)  = \int \varrho _{\mbox{\scriptsize cl}} (p) e ^{ipx / \hbar } dp , \qquad  \rho ^{\mbox{\scriptsize qm}} _{n} (x)  = \int \varrho ^{\mbox{\scriptsize qm}} _{n} (p) e ^{ipx / \hbar } dp , \label{FourierExpansion}
\end{align}
where $\varrho _{\mbox{\scriptsize cl}} (p)$ and $\varrho ^{\mbox{\scriptsize qm}} _{n} (p)$ are the classical and quantum Fourier coefficients, respectively.   Recall that the position and momentum representations of the  wave function are related each other via a Fourier transformation, i.e. $\phi _{n} (p)= \frac{1}{ 2 \pi \hbar } \int \psi _{n} (x)  e ^{- ipx / \hbar } dx$. This implies that the Fourier coefficients in Eq. (\ref{FourierExpansion}) do not correspond to the momentum space probability densities,  e.g.  $\varrho ^{\mbox{\scriptsize qm}} _{n} (p)  \neq \vert \phi _{n} (p) \vert ^{2}$.  Rather, they have the same information content than the probability distributions in position space, just encoded in a different way.

We now take for granted the assumption (iii). Substituting the Fourier expansions (\ref{FourierExpansion}) into Eq. (\ref{LocalAverage}) we obtain
\begin{align}
\int \varrho _{\mbox{\scriptsize cl}} (p) e ^{ipx / \hbar } dp = \lim _{n \gg 1} \int \varrho ^{\mbox{\scriptsize qm}} _{n} (p) \left[ \frac{1}{2 \epsilon _{n}} \int _{x - \epsilon _{n}} ^{x + \epsilon _{n}}  e ^{ipy / \hbar }  \, dy \right] dp  . \label{LocalAverage2}
\end{align}
This integral is quite simple.  Retaining only the leading order term, which is $\epsilon _{n}$-independent, we find
\begin{align}
\int \varrho _{\mbox{\scriptsize cl}} (p) e ^{ipx / \hbar } dp \sim \lim _{n \gg 1} \int \varrho ^{\mbox{\scriptsize qm}} _{n} (p) e ^{ipx / \hbar }  dp  . \label{LocalAverageMomentum}
\end{align}
Finally, due to the linearity of the Fourier expansions (\ref{FourierExpansion}), Eq. (\ref{LocalAverageMomentum}) implies that the quantum Fourier coefficient $\varrho ^{\mbox{\scriptsize qm}} _{n} (p)$ approaches, asymptotically, to the classical Fourier coefficient $\varrho _{\mbox{\scriptsize cl}} (p)$ for $n \gg 1$, i.e.
\begin{align}
\lim _{n \gg 1} \varrho ^{\mbox{\scriptsize qm}} _{n} (p) \sim \varrho _{\mbox{\scriptsize cl}} (p) + \mathcal{O}(1/n)  .  \label{AsymptoticProbDens}
\end{align}
This is an alternative manner to rewrite the local averaging process of Eq.  (\ref{LocalAverage}),  but expressed in the Fourier space.  As evinced by Eq.  (\ref{AsymptoticProbDens}),  the leading term of the asymptotic expansion  corresponds exactly with the classical Fourier coefficient $\varrho _{\mbox{\scriptsize cl}} (p)$,  but subleading terms (in inverse powers of the quantum number $n$) may arise. Evidently, the correction terms vanish when $n$ tends to infinity, thus implying that the quantum Fourier coefficient converge pointwise to its classical analogue. It then seems natural to ask the question, Is the $n \to \infty$ limit a classical limit? We answer this question in the negative by the following reason. Let us consider as a guiding system the quantum harmonic oscillator. 

If an eigenstate is used to evaluate the quantum probability density, in the limit of large quantum number $n$, we obtain strong oscillations and many nodes. As increasing $n$, the quantum distribution confines progressively in a finite region, defined by $\vert x \vert < A _{n}$,  and is small but nonzero for $\vert x \vert > A _{n}$, where $A_{n}=\sqrt{(\hbar / m \omega ) (2n+1)}$.  So, the quantity $A_{n}$ becomes the amplitude of the corresponding classical oscillator.  Since no special assumption concerning the size, mass or $\hbar$ is made,  the limit $n \to \infty$ implies an infinite amplitude, which is nonsense. Thus,  in the classical limit, $n$ has to be large, but finite.  All in all,  our result of Eq. (\ref{AsymptoticProbDens}) suggests that, strictly speaking,  quantum mechanics do not converge to classical mechanics, but rather in the limit of large energy quantum number $n$,  the quantum correction terms will be finite, and they may be interpreted as small quantum corrections at the macroscopic level. We shall now apply this theory to the quantum bouncer, and demonstrate that the leading order term yields the exact classical probability density. The analysis of the quantum corrections is left for a future publication.

\section{Classical limit of the Quantum Bouncer} \label{ClassicalLimit-QB}

The problem of a particle in a constant field force is more than a simple conceptual model. It has found many applications in different branches of physics, for example, it models a particle in a constant electric field, the quantum chromodynamics of quark-antiquark pair at some separation \cite{Charmonium_1978}, and it is often used in modulation doped semiconductors \cite{Weisbuch_1991}. The quantum bouncer is a variation of the constant force problem which adds an infinite wall at the ground level.  It has been used to model the quantum states of neutrons in the Earth's gravitational field \cite{Nesvizhevsky_2002}.  Thus the study of a particle in a constant field should be of wide interest.  We shall now apply the theory developed above to the quantum bouncer. 

We begin our discussion by reviewing briefly the classical problem.  Let us consider a point particle of mass $m$ bouncing on a perfectly reflecting surface, at $z=0$, under the influence of the gravitational acceleration $g$. The corresponding potential energy function is given by
\begin{align}
V (z) = \left\lbrace \begin{array}{c} mgz \\[6pt] \infty \end{array} \quad \begin{array}{l} z \geq 0 \\[6pt] z < 0 \end{array}  \right.  .
\end{align}
Assuming that the particle is initially at rest and dropped from a height $h$,  the classical probability density is easily found to be
\begin{align}
\rho _{\mbox{\scriptsize cl}} (z) = \frac{1}{2 \sqrt{h(h-z)} }  H ( h - z  ) , \label{ClassDensityBouncingBall}
\end{align}
where $H(x)$ is the Heaviside step function. Note that the probability density (\ref{ClassDensityBouncingBall}) does not depend on neither the gravitational acceleration $g$ nor the mass $m$. Besides, it has a divergence at $z=h$, where the particle slows down and reverses direction, but is finite at $z=0$ where the particle simply bounces, reversing direction, but with no change in speed.

The associated quantum mechanical problem, the so called quantum bouncer, is much more difficult. It has been extensively discussed in a number of pedagogical articles and textbooks \cite{Gibbs_1975, Banacloche_1999}.   The normalized wave function $\psi _{n} (z)$ which properly decays for $z \to \infty$ and satisfies the boundary condition $\psi _{n} (z = 0) = 0$ (i.e. the ground is impenetrable), is
\begin{align}
\psi _{n} (z) = \frac{1}{\sqrt{l _{g}}} \frac{ \mbox{Ai} (a _{n} + z / l _{g} ) }{  \mbox{Ai} ^{\prime} (a _{n} ) } , 
\end{align}
where $\mbox{Ai} (x)$ is the Airy function \cite{Vallee_2004}, $a _{n}$ is the $n$-th zero of the Airy function $ \mbox{Ai} (x)$, and $l _{g} = \left( \frac{ \hbar ^{2} }{ 2m ^{2} g } \right) ^{1/3}$ is the gravitational length. Another linearly independent solution to the Schr\"{o}dinger equation exists, the Airy function $ \mbox{Bi} (x)$, which is eliminated as a physical solution since it diverges when the argument becomes large. The corresponding quantized energies $E _{n}$ are 
\begin{align}
E _{n} = - mgl _{g} a _{n} , \qquad n = 1,2,3, \cdots .  \label{EnergySpectrum}
\end{align}
An exact analytical expression for the zeros $a _{n}$ of the Airy function is not available, but a good approximation can be given, especially for large $n$. Using the asymptotic expansion of the Airy function $\mbox{Ai} (x)$ for large negative argument one finds $a _{n} \approx - [(3 \pi / 2 ) ( n - 1/4 ) ] ^{2/3}$ \cite{Vallee_2004}.  This yields an approximate expression for the energy levels (\ref{EnergySpectrum}), which ideed coincides with the WKB quantised spectrum.

\begin{figure}
\includegraphics[scale=0.6]{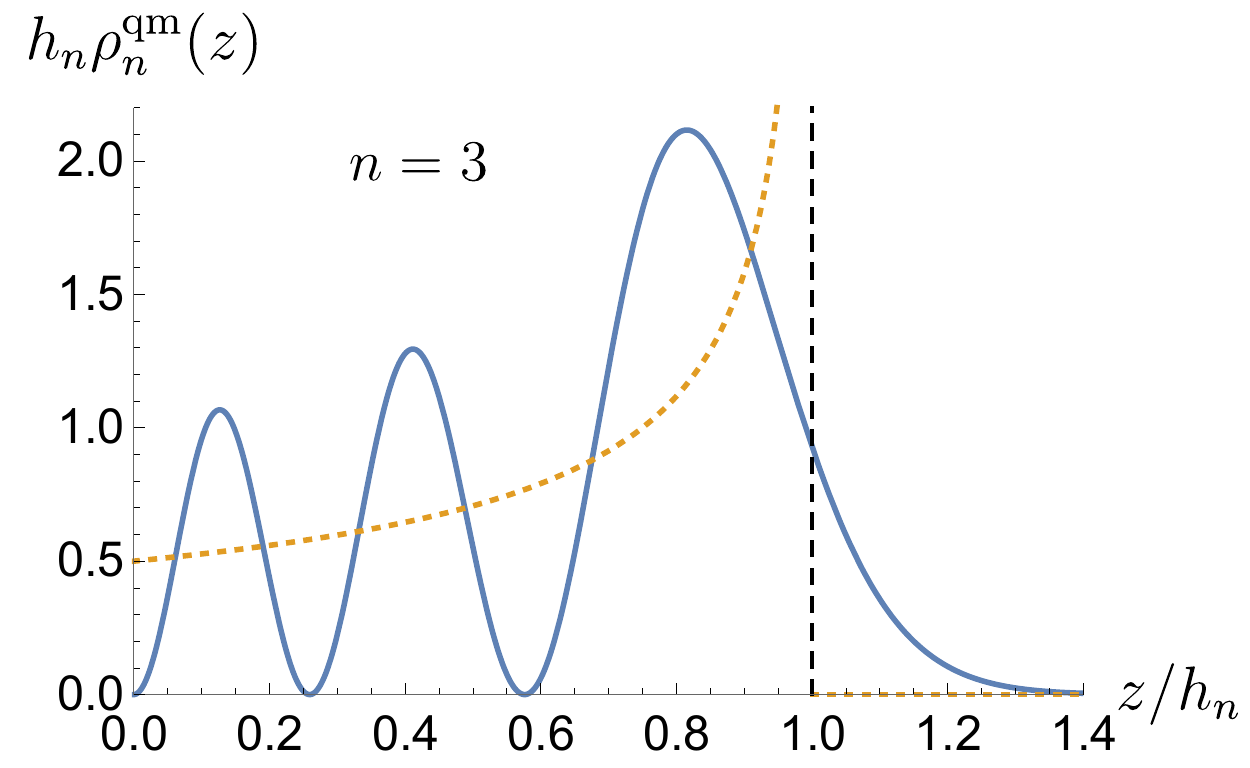} \hspace{1cm}
\includegraphics[scale=0.6]{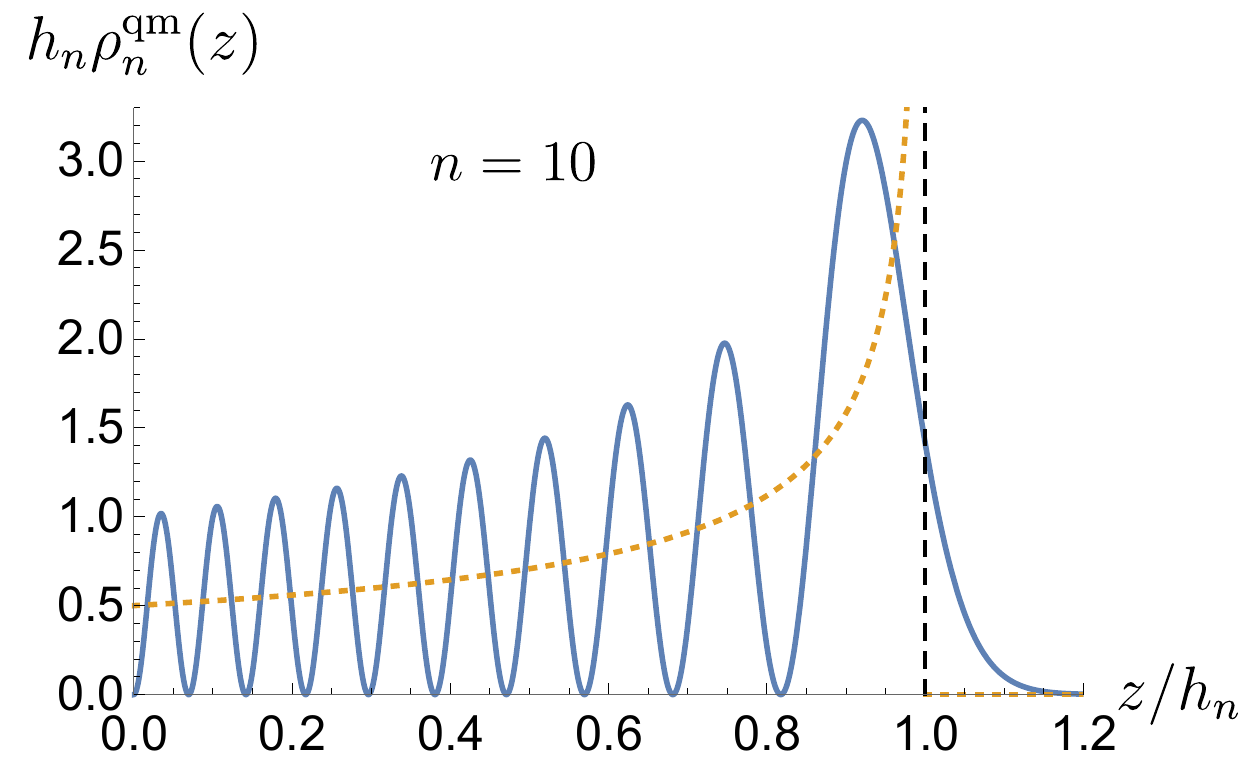} 
\caption{Classical (dashed yellow line) and quantum (continuous blue line) probability densities for energy eigenstates with $n=3$ (at left) and $n=10$ (at right). The dashed vertical line indicates the classical turning point.} \label{QuantumBouncerPlots}
\end{figure}

We are now able to verify visually that the quantum probability density, when locally averaged,  approaches the classical result for large values of $n$.  To this end, we first recall that the quantum distribution exhibits strong oscillations and confines bit by bit between the ground level (at $z=0$) and a finite turning point which depends on the quantum number $n$,  that is determined by equating the expressions for the classical energy and the quantised energy levels.  Within the classical description, a particle with energy $E_{n}$ can rise in the gravitational field up to the height
\begin{align}
h _{n} = - l _{g} a _{n} .   \label{ClassTurningPoint}
\end{align}
The GRANIT experiment has confirmed that a noncoherent beam of ultracold neutrons propagating upwards in the Earth's gravity field reaches precisely the quantized heights (\ref{ClassTurningPoint}) only \cite{Nesvizhevsky_2002}. We now back to the matter.  In figure \ref{QuantumBouncerPlots} we plot the normalised quantum probability density $\rho ^{\mbox{\scriptsize qm}} _{n} (z) = \vert \psi _{n} (z) \vert ^{2}$ (continuous blue line) along with the normalised classical distribution $\rho _{\mbox{\scriptsize cl}} (z)$ (dashed yellow line),  as a function of the normalised distance $z/h _{n}$ and for different values of $n$.  The dashed black vertical line indicates the classical turning point (\ref{ClassTurningPoint}). As shown in the left panel of Fig. \ref{QuantumBouncerPlots},  the quantum probability density for small $n$ has no resemblance with the classical distribution.  However, for large $n$,  as depicted in the right panel of Fig. \ref{QuantumBouncerPlots},  the quantum distribution approaches the classical one in a locally averaged sense, as anticipated. To demonstrate the emergence of the classical behaviour some authors evaluate the probability of finding the particle above the turning point $h_{n}$. A simple calculation yields
\begin{align}
\mathcal{P} _{n} = \int _{h _{n}} ^{\infty} \rho ^{\mbox{\scriptsize qm}} _{n} (z) \, dz =  \left[  \frac{ \mbox{Ai} ^{\prime} (0) }{  \mbox{Ai} ^{\prime} (a _{n}) }  \right] ^{2} ,  \label{ProbabilityBouncer}
\end{align}
implying that, for small quantum numbers, there is a substantial probability that the particle is above the turning point (e.g. $\mathcal{P} _{1}\approx 0.25$, $\mathcal{P} _{2} \approx 0.20$),  whilst it is small for large quantum numbers (e.g. $\mathcal{P} _{10}\approx 0.016$, $\mathcal{P} _{30} \approx 0.0077$).

We now follow the theory developed in the previous section and calculate the classical limit of the quantum bouncer.  We first proceed to the evaluation of the quantum Fourier coefficient $\varrho ^{\mbox{\scriptsize qm}} _{n} (p)$,  defined by Eq. (\ref{FourierExpansion}).  It requires the evaluation of the integral
\begin{align}
\varrho ^{\mbox{\scriptsize qm}} _{n} (p) = \frac{1}{2 \pi \hbar } \frac{1}{l _{g} \mbox{Ai} ^{\prime \, 2} (a _{n} ) } \int _{0} ^{\infty} \mbox{Ai} ^{2} (a _{n} + z / l _{g} ) \, e ^{- ipz / \hbar } dz . \label{FourierCoeffQB}
\end{align}
Up to the best of our knowledge, an exact analytical expression for this integral is still absent in the literature. However, for the purposes of this paper, we only need its asymptotic behaviour for $n \gg 1$.  We achieve this by using the method of Albright \cite{Albright_1977}, since some basic primitives of the Airy functions allow us to express the Fourier coefficient (\ref{FourierCoeffQB}) as a series in inverse powers of $a _{n}$.  To this end,  we first perform the change of variables $x = a _{n} + z / l _{g}$ in Eq.  (\ref{FourierCoeffQB}) and then express the exponential as a power series. The result is
\begin{align}
\varrho ^{\mbox{\scriptsize qm}} _{n} (p) = \frac{1}{2 \pi \hbar } \frac{e ^{ - i p h _{n} / \hbar  } }{ \mbox{Ai} ^{\prime \, 2} (a _{n} ) } \sum _{k=0} ^{\infty} \frac{(- i p l _{g} / \hbar) ^{k}}{k!} \int _{a _{n}} ^{\infty} x ^{k} \mbox{Ai} ^{2} (x)  dx .  \label{FourierCoeffQB2}
\end{align}
The resulting integral can be evaluated by means of the method of Albright \cite{Albright_1977}.  Using the following properties $\mbox{Ai} (a _{n}) = 0$ and  $\mbox{Ai} (x) \to 0$ and $\mbox{Ai} ^{\prime} (x) \to 0$ as $x \to \infty$,  the integral (\ref{FourierCoeffQB2}) becomes  
\begin{align}
\varrho ^{\mbox{\scriptsize qm}} _{n} (p) = \frac{1}{2 \pi \hbar } \frac{e ^{ - i p h _{n} / \hbar  } }{ \mbox{Ai} ^{\prime \, 2} (a _{n} ) } \sum _{k=0} ^{\infty} \frac{(- i p l _{g} / \hbar) ^{k}}{k! } \, \left\lbrace \frac{1}{2k+1} a _{n} ^{k} \mbox{Ai} ^{ \prime \, 2} (a _{n})  +  \frac{k (k-1)(k-2) }{2(2k+1)}    \int _{a _{n}} ^{\infty} x ^{k-3} \, \mbox{Ai} ^{2} (x) \,  dx \right\rbrace .  \label{FourierCoeffQB4}
\end{align}
The resulting integral can be evaluated following the same procedure. Indeed, we can simply iterate the above result. After one iteration we obtain
\begin{align}
\varrho ^{\mbox{\scriptsize qm}} _{n} (p) &= \frac{1}{2 \pi \hbar } \frac{e ^{ - i p h _{n} / \hbar  } }{ \mbox{Ai} ^{\prime \, 2} (a _{n} ) } \sum _{k=0} ^{\infty} \frac{(- i p l _{g} / \hbar) ^{k}}{k! } \, \left\lbrace \frac{1}{2k+1} a _{n} ^{k} \mbox{Ai} ^{ \prime \, 2} (a _{n}) + \frac{k (k-1)(k-2) }{2(2k+1) (2k-5)} a _{n} ^{k-3} \mbox{Ai} ^{ \prime \, 2} (a _{n}) \right. \notag \\ & \hspace{6cm}  \left.   + \frac{k (k-1)(k-2) (k-3) (k-4)(k-5)}{4 (2k+1) (2k-5)}  \int _{a _{n}} ^{\infty} x ^{k-6} \, \mbox{Ai} ^{2} (x) \,  dx \right\rbrace .  \label{FourierCoeffQB5}
\end{align}
We observe that this expression is a series in inverse powers of $a _{n}$.  Recalling that $\vert a _{n} \vert > \vert a_{1} \vert = 2.33811$ for $n>1$, the second term in braces is smaller than the first one by a factor of $a _{n} ^{-3}$, and the third one will be smaller than the second one by an additional factor of $a _{n} ^{-3}$ and so on.  We can perform iterations indefinitely,  however,  for the purposes of this paper,  it is sufficient to keep only the leading terms. In order to justify this assumption,  let us estimate how large the quantum number is in a realistic classical situation. From Eq. (\ref{ClassTurningPoint}),  for a given classical height $h$, it can be estimated as $n \sim \frac{2}{3 \pi} (h / l _{g}) ^{3/2}$,  where we used the asymptotic form of the zeros of the Airy function.
For reference,  for a Caesium atom one finds $l _{g} = 0.226 \mu$m,  such that,  if released from 1mm,  the corresponding quantum number is $n \sim 62,500$ and hence $1/a_{n} ^{3} \sim 10 ^{-11}$.  For the much lighter Na atom one finds $l _{g} = 0.727 \mu$m,  which produces $n \sim 10,000$ so that  $1/a_{n} ^{3} \sim 10 ^{-10}$.  To obtain smaller values for $n$, the gravitational length has to be larger, and this is achieved with light particles.  For example, the corresponding gravitational length for a neutron is $l _{g} = 5.87 \mu$m,  which implies $n \sim 470$ such that $1/a_{n} ^{3} \sim 10 ^{-7}$.  These results imply that we can safely disregard the subleading terms in the asymptotic expansion (\ref{FourierCoeffQB5}), and retain only the leading term, i.e.
\begin{align}
\varrho ^{\mbox{\scriptsize qm}} _{n} (p) \sim \frac{1}{2 \pi \hbar } e ^{ - i Q }  \sum _{k=0} ^{\infty} \frac{( i Q ) ^{k}}{k! (2k+1)} ,  \label{FourierCoeffQB6}
\end{align}
where $Q \equiv p h _{n} / \hbar$.  This summation can be performed in an analytical fashion. The result is
\begin{align}
\varrho ^{\mbox{\scriptsize qm}} _{n} (p) \sim  \frac{1}{4 \hbar \sqrt{\pi Q} } \,  e ^{ - i ( Q - \pi / 4) } \,  \text{erf} \left[ e ^{- i \pi / 4} \sqrt{Q} \right]  , \label{FourierCoeffQBZeroth}
\end{align}
where $\text{erf} \, (x)$ is the error function,  defined as $\text{erf} \, (x) = \frac{2}{\sqrt{\pi}} \int _{0} ^{x} e ^{-t ^{2}} dt$,  that in turn can also be expressed in terms of the Fresnel integrals \cite{Gradshteyn_Ryzhik}.  The summations of the subdominant terms can also be performed analytically, they are however not relevant for the purposes of this paper.

According to the theory developed in the previous section,  to obtain the asymptotic behaviour of the quantum probability density in position space,  we have to compute the Fourier transformation of the coefficient $\varrho ^{\mbox{\scriptsize qm}} _{n} (p)$,  given by Eq. (\ref{FourierCoeffQBZeroth}),  as defined in Eq. (\ref{FourierExpansion}).  This yields
\begin{align}
\rho ^{\mbox{\scriptsize qm}} _{n} (z) & \sim \frac{1}{4 \sqrt{\pi} h _{n} } e ^{ i \pi / 4} \int _{- \infty} ^{\infty} \text{erf} \left[ e ^{- i \pi / 4} \sqrt{Q} \right] \, e ^{ - i Q (1 - z/ h _{n} ) }  \frac{dQ}{\sqrt{Q}} =  \frac{1}{2 \sqrt{h_{n}(h_{n}-z)} }  H ( h_{n} - z  )  , 
\end{align}
which is exactly the classical probability density, $\rho _{\mbox{\scriptsize cl}} (z)$, given by Eq. (\ref{ClassDensityBouncingBall}),  with the classical height $h$ replaced by the turning point $h _{n}$ of Eq.   (\ref{ClassTurningPoint}).  Following the same procedure, in Refs.  \cite{ClassLim1,ClassLim3,ClassLim2} we successfully applied this method to other periodic systems, such as the harmonic oscillator, the infinite square well potential and the hydrogen atom.

We close this section by commenting some previous works related with the classical limit of the quantum bouncer.  Many pedagogical articles show the comparison between the classical and quantum distributions for this system, asserting that they approach each other in a locally averaged sense when the energy quantum number is large. However, none of them have computed the local averages to confirm this claim, either analytically nor numerically.  In a recent paper \cite{Singh_2016},  the authors consider a particle in a homogeneous field and analyse the expectation values of the position, momentum, and their moments up to fourth order (without restoring to numerical or graphical techniques), to confirm that they reduce to the corresponding classical values for large quantum numbers.  Note that all the results reported in Ref.  \cite{Singh_2016},  are just particular cases of our findings in this paper. For example, having showed that the quantum distribution $\rho ^{\mbox{\scriptsize qm}} _{n} (z)$ converges to its classical analogue $\rho _{\mbox{\scriptsize cl}} (z)$ for large quantum number $n \gg 1$, then the expectation value of any observable $f$ depending on the position $z$ correctly yields the classical result:
\begin{align}
\braket{f(z)} _{\mbox{\scriptsize qm}} = \int _{0} ^{\infty} dz \, f (z) \rho ^{\mbox{\scriptsize qm}} _{n} (z) \; \sim \; \int _{0} ^{\infty} dz \, f (z) \rho _{\mbox{\scriptsize cl}} (z) = \braket{f(z)} _{\mbox{\scriptsize cl}} . \label{PosExpectationValue}
\end{align}
In a similar fashion,  the expectation value of any observable $g$ depending on the momentum $p$,  reduces to the classical result.  To demonstrate this claim, we consider that the observable $g$ can be written as a power series, i.e. $g(p)= \sum _{k = 0} ^{\infty} c_{k} p ^{k}$.  Because of parity considerations, the expectation value of the odd terms vanish, so we are left only the the even terms in the power expansion.  However,  using the quantum Hamiltonian $H = \frac{p^{2}}{2m} + mgz$, it is clear that
\begin{align}
\int _{0} ^{\infty} dz \,  \psi _{n} ^{\ast} (z) p ^{2k} \psi _{n} (z) = \int _{0} ^{\infty} dz \,  \left[ 2m (H - mgz ) \right] ^{k}  \rho ^{\mbox{\scriptsize qm}} _{n} (z) , \label{MonExpectationValue}
\end{align}
which is simply evaluated by using the result of Eq.  (\ref{PosExpectationValue}). Therefore, all the results reported in Ref.  \cite{Singh_2016} are easily reproduced with our method.

Finally,  in Ref.  \cite{Leubner_1988} the authors pointed out that comparison between classical and quantum probability densities is unwarranted when applied to pure energy eigenstates. Instead, they claim that it should be applied to a mixed ensemble of quantum coherent states, and hence they compare the quantum distribution with an ensemble of classical systems (each with the same initial conditions).  A similar analysis was undertaken in Ref.  \cite{Cabrera_1987},  where the authors investigate the large quantum number behaviour of a quantum state which consists of a superposition of a few successive energy eigenstates for the harmonic oscillator.  Their conclusions are the same: quantum effects persist for arbitrarily large quantum number,  and they do not recover the exact classical distribution.  There are important conceptual differences between these approaches and the one introduced in this paper: while they consider an ensemble of classical systems \cite{Leubner_1988} or a superposition of energy eigenstates \cite{Cabrera_1987}  here we apply the correspondence principle to a single-particle energy eigenstate, and obtain the exact classical result in an analytical fashion.

\section{Discussion and outlook} \label{Conclusions}


Our results have interesting implications in the understanding of the classical limit.  To better understand the scope and limitations of our approach, we shall discuss briefly various formulations of the classical limit, by critically reviewing their main advantages and drawbacks. As discussed in Sec. \ref{Intro}, there are two different formulations of the correspondence principle.  The first is traced back to Planck, in 1906, which asserted that classical physics is recovered from quantum physics in the limit $h \to 0$.  Planck originally introduced this limit to show that his energy density for blackbody radiation,
\begin{align}
\rho (\nu , T ) = \frac{8 \pi h }{c ^{3}} \frac{\nu ^{3}}{e ^{ h \nu / k _{\mbox{\scriptsize B}} T} - 1} ,
\end{align} 
reduces in the limit $h \to 0$ to the classical Rayleigh-Jeans energy density, 
\begin{align}
\rho _{\mbox{\scriptsize cl}} (\nu , T ) = \lim _{h \to 0 } \rho (\nu , T ) = \frac{8 \pi \nu ^{2} }{c ^{3}} k _{\mbox{\scriptsize B}} T .
\end{align}
In these expressions $ k _{\mbox{\scriptsize B}}$ is the Boltzmann constant and $T$ is the temperature. Clearly,  Planck applied the correspondence principle to a statistical average of the energy eigenvalues over a canonical probability distribution, without appealing to any quantum number.  Years later, in 1913, Bohr postulated that quantum mechanics reduces to classical mechanics when the energy quantum number is large, $n \gg 1 $. Bohr first enunciated this formulation to show that in his model of the hydrogen atom the transition frequency between neighbouring energy levels,
\begin{align}
\nu = \frac{me ^{4}}{64 \pi ^{3} \epsilon _{0} ^{2} \hbar ^{3}} \left[ \frac{1}{n^{2}} - \frac{1}{(n+1)^{2}} \right] ,
\end{align}
in the limit of large quantum numbers $n \gg 1$, tends to the classical orbital frequency of the electron
\begin{align}
 \nu _{\mbox{\scriptsize cl}} = \frac{1}{2 \pi} \sqrt{\frac{e ^{2}}{4 \pi \epsilon _{0}  m r ^{3} _{n} }} ,
\end{align}
where $r _{n} = n ^{2} r _{\mbox{\scriptsize B}}$ is the radius of stationary orbits and $r _{\mbox{\scriptsize B}}$ is the Bohr radius. So, Bohr applied his correspondence principle to a single atom, not to an statistical ensemble of them. The inadequacy of Bohr's correspondence principle to connect quantum and classical theory has been extensively discussed in Refs. \cite{Gao_1999, Eltschka_2001, Boisseau_2001}. Also, as shown in Refs.  \cite{Liboff_1975, Liboff_1979}, Planck and Bohr correspondence principles are not universally equivalent when applied to transition frequencies. This is clearly seen for a particle in a box, for which the transition frequency between adjacent energy levels is $\nu = \frac{h}{8 m d ^{2}}  (2n-1)$, 
such that the separate limits $h \to 0$ and $n \to \infty$ give zero and infinity, respectively.  However, as suggested in Ref.  \cite{Hassoun_Kobe_1989}, taking simultaneously the limits $h \to 0$ and $n \to \infty$ but keeping finite the classical action $J=nh$, the classical frequency is exactly recovered. 

It is worth mentioning that both Planck and Bohr formulations of the correspondence principle were introduced before the development of wave mechanics by Schr\"{o}dinger and matrix mechanics by Heisenberg, Born and Jordan in 1925, and so they did not make any allusion to the classical limit of quantum dynamics. Therefore, it was not clear how to apply these correspondence principles to whether the wave function or the matrix elements of an operator. Along the years, a few methods addressing the quantum-to-classical correspondence have been proposed within modern quantum mechanics, all of them rooted in the old Planck and Bohr correspondence principles, namely, the WKB method, the path integral formulation and the Ehrenfest's theorem. As we shall discuss in the following, these methods are not universally reliable for investigating the classical limit of quantum dynamics in a general way, and at the end all of them require of a very narrowly peaked probability density.

\begin{itemize}

\item The WKB method, also called the semiclassical approximation, is important as a practical means of approximating solutions to the Schr\"{o}dinger equation for slowly varying potentials compared to the de Broglie wavelength $\lambda = h/p$ of the particle \cite{Wentzel_1926, Kramers_1926, Brillouin_1926}.  The basic idea is that the wavefunction can be written as the exponential of a complex function $\mathcal{S}(x)$, i.e. $\psi _{\mbox{\scriptsize WKB}} (x) = e ^{i \mathcal{S}(x) / \hbar}$. Plugging this ansatz into the Schr\"{o}dinger equation leaves us with the nonlinear differential equation
\begin{align}
i \hbar \frac{\partial ^{2} \mathcal{S} }{ \partial x ^{2}} - \left( \frac{\partial \mathcal{S} }{ \partial x } \right) ^{2} + p ^{2} (x) =0 , \label{WKB}
\end{align}
where $p (x) = \sqrt{2m [E - V(x)]}$ is the momentum.  Inserting the so-called WKB expansion $\mathcal{S} = S + (\hbar / i ) S _{1} + (\hbar / i ) ^{2} S _{2} + \cdots $ into Eq. (\ref{WKB}) we then find that the function $S$ satisfies the Hamilton-Jacobi equation,
whose solution is the Hamilton's principal function $S(x)= \pm \int ^{x}  p (x ^{\prime}) dx ^{\prime}$. At $\mathcal{O} (\hbar)$ one finds $S_{1} (x) = - \tfrac{1}{2} \ln p (x)$.  Putting these together gives us the WKB wave function,
\begin{align}
\psi _{\mbox{\scriptsize WKB}} (x) \sim \frac{1}{\sqrt{p(x)}} \exp \left[ \pm \frac{i}{\hbar} \int ^{x}  p (x ^{\prime}) dx ^{\prime} \right] , \label{WKB-WaveFunction}
\end{align}
Clearly, the WKB wave function (\ref{WKB-WaveFunction}) diverges near the turning points of the classical motion, where the velocity component vanishes. Even though the WKB method sheeds light about the classical limit of the Schr\"{o}dinger equation, it suffers from some difficulties, that indicate its lack of generality for calculating the classical limit of quantum mechanics. For example,  superposition of WKB wave functions, such as $\psi (x) = e ^{i \mathcal{S}(x) / \hbar} + e ^{- i \mathcal{S}(x) / \hbar}$, does not lead to the classical Hamilton-Jacobi equation.  Besides,  there are physical systems which cannot be described by WKB wave functions. For example, it is not adequate to describe the circular orbits in the hydrogen atom, since the electron is always at a turning point, where the WKB method breaks down \cite{Sengupta_2004,  Sengupta_2005,  Sengupta_2006}. This would mean that the circular orbits in the hydrogen atom do not have classical limit, which is clearly nonsense.

\item Feynman path integral formulation of nonrelativistic quantum mechanics provides a different conceptual framework which is completely equivalent to that given by Schr\"{o}dinger's equation. It passes from standard  Hamiltonian formulation to a Lagrangian description in which there are only commuting objects, not operators.  In this approach, the quantum nature arises because to compute a quantum amplitude, one considers not a unique classical trajectory, but a sum (or functional integral) over an infinity of possible trajectories. 
The contribution of a path is proportional to $e ^{i S [x(t)] / \hbar }$, where $S [x(t)] = \int _{x(t)} \mathcal{L} (x, \dot{x}) dt$ is the action given by the time integral of the Lagrangian $\mathcal{L} (x, \dot{x})$ along the path $x(t)$.  If $S \gg h$,  the exponential in the Feynman's kernel
\begin{align}
K (x,t:x^{\prime},t^{\prime}) \sim \int _{x^{\prime},t^{\prime}}^{x,t} d[x(t)] \,  \exp \left\lbrace \frac{i}{\hbar} S [x(t)] \right\rbrace \label{propagator}
\end{align}
oscillates rapidly and the path integral will be dominated by contributions from the extrema of the action (by the usual stationary phase approximation), i.e. precisely by those paths that are solutions to the Lagrange's equation \cite{FeynmanHibbs_1965}. However this cannot be regarded as a complete description of the classical limit. To get a full classical limit, one needs to specify the initial and final positions of the particle. In the quantum formalism instead we are able to provide only the amplitudes for a particle to be in a given position at a given time, and at best a sharply peaked probability density with well-defined position and momentum. Consequently, although the Planck's limit correctly gives to the Lagrange's equations in the path integral formalism, it does not say how to obtain the required sharply peaked distribution to classically define the initial and final points.

\item Many textbooks discuss the classical limit in terms of Ehrenfest's theorem. The essential points of the theorem can be illustrated for a particle moving in a one-dimensional potential $V(x)$.  The Heisenberg equations of motion for position and momentum, averaged in some quantum state,  are
\begin{align}
\frac{d \braket{p}}{dt} = \braket{ F(x) } , \qquad \frac{d \braket{x}}{dt} = \frac{\braket{p}}{m} , \label{Ehrenfest}
\end{align}
where $F(x) = -  \partial V (x) / \partial x$ is the force operator. At first glance, it might appear that Ehrenfest's theorem says that the mean position and momentum obey Hamilton's equations of motion, but this is not actually the case. If they were to satisfy Hamilton's equations, the right-hand side of the first equation would have to be $F( \braket{ x } )$ instead of $\braket{ F(x) }$, which are quite different in general. Consequently, Ehrenfest's theorem does not say that the quantum motion follows a classical trajectory. At best, it ensures that if the uncertainties $\Delta x$ and $\Delta p$ are both sufficiently small, the quantum motion will approximate the classical trajectory. 
Corrections to Ehrenfest's theorem can be derived as follows.  Taylor expanding the force operator $F(x)$ around the mean position $\bar{x} = \braket{x}$ one gets \cite{Ballentine_1994}
\begin{align}
\frac{d \bar{p} }{dt} \approx F( \bar{x} ) + \frac{1}{2} \braket{(\delta x) ^{2}} \frac{\partial ^{2} F( \bar{x} )}{\partial \bar{x} ^{2}} + \cdots   , \label{Ehrenfest2}
\end{align}
where $\bar{p} = \braket{p}$ is the mean momentum and $\delta x = x - \bar{x}$ is an operator for the deviation from the mean value.  Clearly, $\bar{x}$ and $\bar{p}$ obey the classical equations only if the correction terms in Eq. (\ref{Ehrenfest2}) are negligible, for which a very narrowly peaked probability density is needed. Therefore, although Ehrenfest's theorem does not resort directly to Planck or Bohr correspondence principles,  one needs to appeal to some of them to get the required sharp narrow probability density. All in all, Ehrenfest's theorem can not be regarded as a general description of the classical limit of the quantum theory.

\end{itemize}

From the above we have learned that the probability density plays a fundamental role to the fulfilment of the classical limit within the different approaches. However, none of them tackles the problem of how a narrowly peaked distribution can be obtained from the quantum one. In this regard, many textbooks and articles make direct comparisons between the quantum and classical probability densities for periodic systems.  The former, defined by $\rho ^{\mbox{\scriptsize qm}} _{n} (x) = \vert \psi _{n} (x) \vert ^{2}$, is intrinsic within quantum mechanics, while the latter, given by $\rho _{\mbox{\scriptsize cl}} (x) = \frac{2}{\tau \vert v (x) \vert }$, is intuitively derived by the argument of how much time spent the particle in a given interval.  These quantities are fundamentally different: whilst the quantum distribution is defined in the whole real line and is oscillatory, the classical distribution is a smooth function defined between the turning points.  Therefore, it is clear that $\rho _{\mbox{\scriptsize cl}} (x)$ cannot be obtained from $\rho ^{\mbox{\scriptsize qm}} _{n} (x)$ by taking directly the Bohr's limit $n \gg 1$. However what is relevant in the comparison between the distributions is that, for a large energy quantum number, the quantum distribution follows its classical counterpart in a locally averaged sense, as defined by Eq. (\ref{LocalAverage}). In principle, one could use the WKB wave function to do this, but it diverges near the turning points. An exact wave function instead does not actually diverge there, but become sharply peaked around, thus resembling to the classical turning points, but smoothed out by the uncertainty principle, which also causes some of the wave function to spill over into the classically forbidden region. In general, local averaging exact quantum distributions in position space is a difficult task due to its highly oscillatory behaviour.  To fill in this gap,  in this paper we develop a systematic approach to determine the classical limit of periodic quantum systems.

Our method lies in three simple assumptions: (i) the classical motion is periodic, (ii) the quantum probability density rapidly oscillates for large quantum numbers, and (iii) the classical and quantum distributions approach each other for large $n$. Assumptions (i) and (ii) allow us to express the quantum probability density as a Fourier expansion (\ref{FourierExpansion}), and the latter (iii) implies that, for large quantum numbers, the Fourier coefficient converges to its classical counterpart (\ref{AsymptoticProbDens}). In fact, the classical Fourier coefficient is just the leading term in the asymptotic expansion of the quantum Fourier coefficient, and hence, quantum corrections may arise in inverse powers of the quantum number $n$ (or equivalently in powers of the WKB ratio $h/S \ll 1$, where $S$ is the classical action).  Clearly, we assume that both $h$ and $n$,  have to be finite, which is consistent with the fact that Planck constant is a nonzero universal constant of nature (and is unwarranted to make it exactly zero) and that the energy of a classical system, although large as compared with the quantum of energy,  must be finite.


Following our method, we obtain the exact classical limit of a particle bouncing in the presence of the gravitational field.  In this case,  the quantum probability distribution is related to the Airy function,  so the computation of the exact Fourier coefficient can be treated numerically, at best. However, our method requires only its asymptotic behaviour for large quantum number, which can be obtained analytically by means of the method of Albright, as showed. As expected, we obtain a series in inverse powers of $a_n$, the zeros of the Airy function. Our estimations showed that for large $n$, the first subdominant term is strongly suppressed with respect to the leading term. Retaining only the dominant term and going back again to the position space (through the inverse Fourier transformation), we obtain the exact classical probability density. This is precisely the main result of this paper. In a forthcoming publication we will report the analysis of the quantum corrections, which as discussed, are small at the macroscopic level but nonzero. Indeed, they consist of small oscillations enveloping the classical distribution, whose amplitude is of the order of $a_{n} ^{-3}$  \cite{Bernal_2022}. Physically,  this seems that, by a conspiracy of nature, the quantum nature of the world is hidden at the macroscopic level, leaving us with an apparent world described consistently by classical physics. 

Finally, we suggest some possible avenues of research and open problems, amenable to explore.  First, we encourage the calculation of the exact classical limit of other well-known periodic quantum systems in a given energy eigenstate.  Two of us have applied this method to the harmonic oscillator, the infinite square well potential and the hydrogen atom, and showed that it gives the exact classical probability distributions. Also, we suggest the analysis of the classical limit of quantum systems described by a superposition of energy eigenstates. We expect an statistical answer like this: the classical probability must correspond to the diagonal terms in the density matrix, while nondiagonal terms should be strongly suppressed in the high energy limit, according to our classical experience. It would be interesting also to analyse the inclusion of the spin in the classical limit procedure, does it make sense? These problems are to be solved in the near future.

\appendix

\section*{Acknowledgements}
A.M.-R. has been partially supported by DGAPA-UNAM Project No. IA102722 and by Project CONACyT (M\'{e}xico) No. 428214.


\bibliography{Bib.bib}

\end{document}